*Beamforming Techniques for Multichannel audio Signal Separation*

[1]Hidri Adel, [2]Meddeb Souad, [3]Abdulqadir Alaqeeli, [4]Amiri Hamid
*1, Corresponding author National Engineering School of Tunis, hidri_adel@yahoo.fr*
*[2,4] National Engineering School of Tunis, mmemeddeb@gmail.com, hamidlamiri@yahoo.com*
*[3]King Abdul Aziz City for Science and Technology, aalaqeeli@kacst.edu.sa*

*Abstract*

*Beamforming is a signal processing technique. It has been studied in many areas such as radar, sonar, seismology and wireless communications, to name but a few. It can be used for a myriad of purposes, such as detecting the presence of a signal, estimating the direction of arrival, and enhancing a desired signal from its measurements corrupted by noise, competing sources and reverberation. Actually, Beamforming has been adopted by the audio research society, mostly to separate or extract speech for noisy environment. Beamforming techniques basically approach the problem from a spatial point of view. A microphone array is used to form a spatial filter which can extract a signal from a specific direction and reduce the contamination of signals from other directions. In this paper we survey some Beamforming techniques used for multichannel audio signal separation.*

**Keywords**: *Beamforming, Microphone array, Signal Separation, Deterministic Beamformer, Statistically optimum Beamformer.*

## 1. Introduction

Most speech signal receiving systems operate in noisy environments, where the desired speech signal is corrupted by interfering signals such as competing speakers and noise sources, and also distorted by the reverberating environment. In many applications, there is a need to separate the multiple sources or extract a source of interest while minimizing undesired interfering signals and noise. The estimated signals may then be either directly listened to or further processed [1].

It is proven scientifically that humans are able to separate only one conversation in a highly noisy environment, such as in a cocktail party environment. Inspite of being studied for decades, the cocktail party problem remains a scientific challenge that demands further research efforts [2] [3].

As foregrounded in some recent works [4], using a single channel it is not possible to improve both intelligibility and quality of the recovered signal at the same time. A way to overcome this limitation is to add some spatial information to the time/frequency information available in the single channel case. We can get this additional information using two or more channel of noisy speech named multichannel.

There are two main categories of multichannel (microphone array) algorithms, namely: Blind Source Separation (BSS) and Beamforming. BSS is an approach for estimating source signals using only information about their mixtures observed in each input channel. The estimation is performed without possessing information on each source, such as its frequency characteristics and location, or on how the sources are mixed. On the other hand, the Beamforming family of algorithms concentrates on enhancing the sum of the desired sources while treating all other signals as interfering sources. Since the BSS family of algorithms is not focal point of this paper, it will focus on the Beamforming techniques.

A Beamformer is a signal processor used together with a microphone array to provide the capability of spatial filtering. The microphone array produces spatial samples of the propagating wave, which are then manipulated by the signal processor to produce the Beamformer output signal. Beamforming is accomplished by filtering the microphone signals and combining the outputs to extract (by constructive combining) the desired signal and reject (by destructive combining) interfering signals according to their spatial location. Beamforming can separate sources with overlapping frequency content that originate at different spatial locations [1,5].

Speech separation based on Beamforming techniques have been intensively studied in recent years due to their many applications [6, 7, 8, 9]. These techniques can be classified into two categories, depending on the approach taken to estimate the spatial filter weights: deterministic Beamforming approaches and statistically optimum one.

The rest of the paper is organized as follow: section 2 presents the problem definition. Section 3 gives some Beamforming techniques. Section 4 discuss presented techniques and states the problem. Section 5 concludes the paper and gives future work.

## 2. Problem definition

Audio mixtures can be reached through many ways. This results in different characteristics of the sources and the mixing process that can be exploited by the separation methods. The observed spatial properties of audio signals depend on the spatial distribution of a sound source, the sound scene acoustics, the distance between the source and the microphones, and the directivity of the microphones.

In General, the problem of Multichannel Source Separation (MSS) is stated to be the process of estimating the signals from *N* unobserved sources, given from *M* microphones, which arises when the signals from the *N* unobserved sources are linearly mixed together as presented in figure 1.

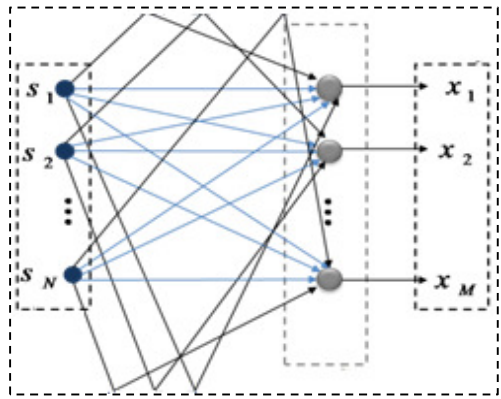

**Figure1.** Multichannel problem formulation

The signal recorded at the $j^{th}$ microphone can be modeled as:

$$x_j(n) = \sum_{i=1}^{N} \sum_{p=1}^{P} h_{ji}^p(p) \, S_i\left(n - \Delta_{ji}^p\right) \quad j = 1 \ldots M \tag{1}$$

Where $S_i$ and $x_j$ are the source and mixture signals respectively, $h_{ji}$ is a P-point Room Impulse Response (RIR) from source *i* to microphone *j*, *P* is the number of paths between each source-microphone pair and $\Delta$ is the delay of the $p^{th}$ path from source *j* to microphone *i* [6, 7]. This model is the most natural mixing model, encountered in live recordings called echoic mixtures.

In reverberation free environments (*P=1*), the samples of each source signal can arrive at the microphones only from the line of sight path, and the attenuation and delay of source *i* would be determined by the physical position of the source relative to the microphones. This model, named anechoic mixing, is described by the following equation derived from the previous equation:

$$x_j(n) = \sum_{i=1}^{N} h_{ji} \, S_i(n - \Delta_{ji}), \quad j = 1..M \tag{2}$$

The instantaneous mixing model is a specific case of the anechoic mixing model where the samples of each source arrive at the microphones at the same time $(\Delta_{ji} = 0)$ with differing attenuations, each elements of the mixing matrix $h_{ji}$ is a scalar that represents the amplitude scaling between source *i* and microphone *j*. From the equation (2), instantaneous mixing model can be expressed as:

$$x_j(n) = \sum_{i=1}^{N} h_{ji} S_i(n), \qquad j = 1..M \qquad (3)$$

## 3. Beamforming Techniques

Beamforming is a class of algorithms for multichannel signal processing. The term Beamforming refers to the design of a spatio-temporal filter which operates on the outputs of the microphone array [10]. This spatial filter can be viewed in terms of dependence upon angle and frequency. Beamforming is achieved by filtering the microphone signals and combining the outputs to extract (by constructive combining) the desired signal and reject (by destructive combining) interfering signals according to their spatial location.

Beamforming for broadband signals like speech can, in general, be performed in the time domain or frequency domain. In time domain Beamforming, a Finite Impulse Response (FIR) filter is applied to each microphone signal, and the filter outputs combined to form the Beamformer output. Beamforming can be performed by computing multichannel filters whose output is $\hat{s}(t)$ an estimate of the desired source signal. The output can be expressed as:

$$\hat{s}(t) = \sum_{i=1}^{N} \sum_{p=0}^{P-1} w_{i,p} x_i(t-p) \qquad (4)$$

Where P-1 is the number of delays in each of the N filters. In frequency domain Beamforming, the microphone signal is separated into narrowband frequency bins using a STFT, and the data in each frequency bin is processed separately. Beamforming techniques can be broadly classified as being either data-independent, or data-dependent. Data-independent, or deterministic, Beamformers are so named because their filters do not depend on the microphone signals and are chosen to approximate a desired response. Conversely, data-dependent or statistically optimum Beamforming techniques, their filters are designed based on the statistics of the arriving data to optimize some function that makes the Beamformer optimum in some sense.

### 3.1. Deterministic Beamformer

The filters in a deterministic Beamformer do not depend on the microphone signals and are chosen to approximate a desired response. For example, one may wish to receive any signal caming from a certain direction, in which case the desired response is unity over at that direction. As another example, we may know that there is interference operating at a certain frequency and arriving from a certain direction, in which case the desired response at that frequency and direction is zero. The simplest deterministic Beamforming technique is delay-and-sum Beamforming, where the signals at the microphones are delayed and then summed in order to combine the signal arriving from the direction of the desired source coherently, expecting that the interference components arriving from off the desired direction cancel to a certain extent by destructive combining. The delay-and-sum Beamformer as shown in Figure 2 is simple in its implementation and provides for easy steering of the beam towards the desired source. Assuming that the broadband signal can be decomposed into narrowband frequency bins, the delays can be approximated by phase shifts in each frequency band.

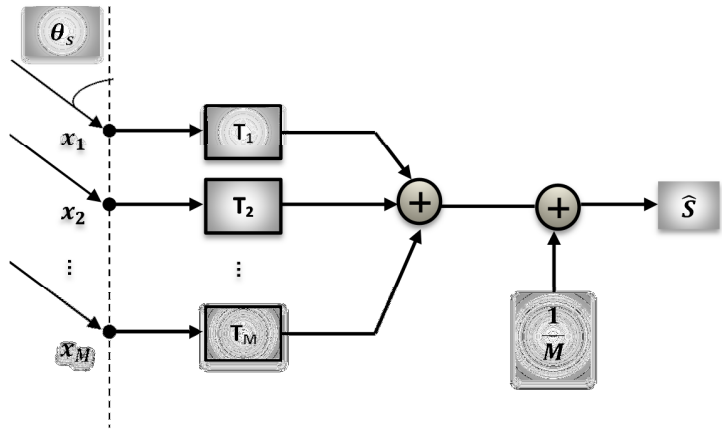

**Figure 2.** Delay-and-sum Beamforming

The performance of the delay-and-sum Beamformer in reverberant environments is often insufficient. A more general processing model is the filter-and-sum Beamformer as shown in Figure 3

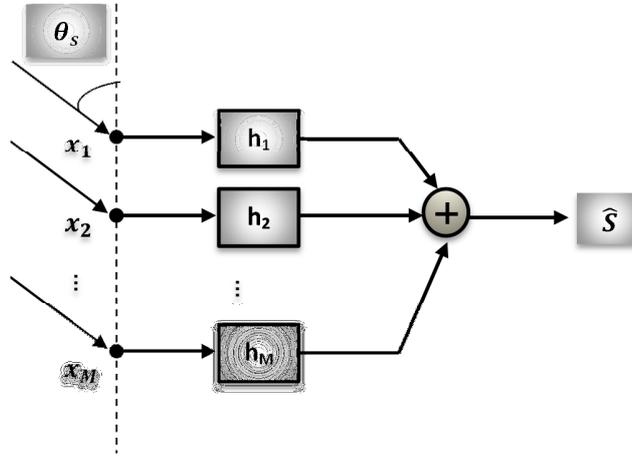

**Figure 3.** Filter and sum Beamforming

Where, before summation, each microphone signal is filtered with FIR filter of order M. This structure, designed for multipath environments, namely reverberant enclosures, replaces the simpler delay compensator with a matched filter.

### 3.2. Statistically optimum Beamformer

Statistically optimal Beamformers are designed based on the statistical properties of the desired and interference signals. In this category, the filters are designed based on the statistics of the arriving data to optimize some function that makes the Beamformer optimum in some sense. Several criteria can be applied in the design of the Beamformer, e.g., maximum signal-to-noise ratio (MSNR), minimum mean-squared error (MMSE), minimum variance distortionless response (MVDR) and linear constraint minimum variance (LCMV). A summary of several design criteria can be found in [11]. In general, their goal is to optimize the Beamformer response so the output contains minimal contributions due to noise and interfering signals.

If we consider the performance criteria is to minimize the Mean Square Error between the Beamformer output and the desired source. The cost function to be minimized is given as:

$$J = E[|d(n) - y(n)|^2] \tag{5}$$

Where E [.] denotes the expectation operator (or ensemble average). Substituting for the output of the Beamformer as $y(n) = w^H x(n)$ and taking of the gradient of the cost function and setting it to zero, we get:

$$\nabla J = -2r_{xd} + 2R_{xx}w = 0 \quad (6)$$

Where $R_{xx} = E[x(n)x^H(n)]$ is the correlation matrix of the input signal $x(n)$. Also $r_{xd} = [x(n)d^*(n)]$ is the cross-correlation vector between the sensor inputs and the desired signal $d(n)$.

Solving equation (6) gives the expression for the optimum weights for MMSE as:

$$w_{MMSE} = R_{xx}^{-1} r_{xd} \quad (7)$$

The aforementioned equation is also known as the Wiener-Hopf equation or the Optimum Wiener solution. This Beamformer can be viewed as a multichannel Wiener filter (MWF). In [12], a MWF technique was proposed. The MWF produces an MMSE estimate of the desired speech component in one of the microphone signals, hence simultaneously performing noise reduction and limiting speech distortion. In addition, the MWF is able to take speech distortion into account in its optimization criterion, resulting in the speech distortion weighted multichannel Wiener filter (SDW-MWF) [13]. One method of improving the system performance is to add a post-filter to the output of the Beamformer as shown in the following figure:

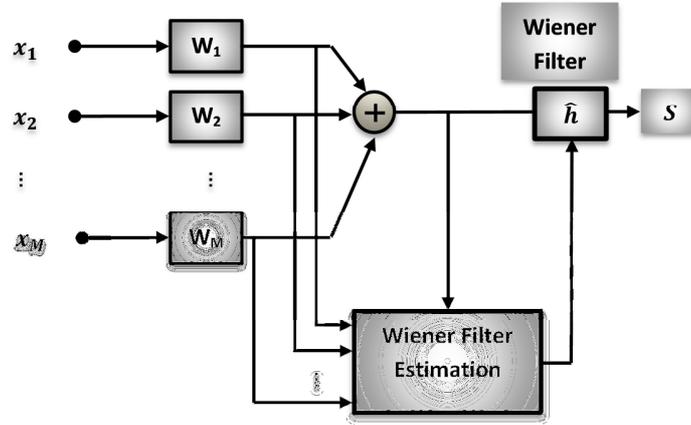

**Figure 4.** Filter and sum Beamformer with post-filter

The statistics of the array data are not usually known and may change over time so adaptive algorithms are typically used to determine the weights. The adaptive algorithm is designed so the Beamformer response converges to a statistically optimum solution.

Adaptive algorithms determine the weights iteratively approximating the optimum weights. The reason being that computing optimum weights using equation (7) involves ensemble averaging and matrix inversions. The most common and widely used adaptive algorithms are Least Mean-Square (LMS) and Recursive Least Squares (RLS).

The LMS algorithm is an MMSE weight adaptation algorithm that uses the steepest descent algorithm. The weight vectors are calculated recursively using the following equations:

$$w(k+1) = w(k) + \Delta x(n)e(n), \quad e(n) = d^*(n) - x^H(n)w(n) \quad (10)$$

In the above equation, $\Delta$ is the step-size and determines the rate of convergence of the algorithm. The choice of $\Delta$ depends on the eigen-spread of the covariance matrix $R_{xx}$.

The LMS algorithm requires knowledge of the transmitted signal. This is usually accomplished by sending periodically some known pilot sequences that is usually known to the receiver.

The RLS algorithm uses weighted sums for estimating $R_{xx}$ and $r_{xd}$ using the following equations:

$$\bar{R}_{xx} = \sum_i \gamma^{n-1}[x(i)x^H(i)], \quad \bar{r}_{xd} = \sum_i \gamma^{n-1}d^*(i)x(i) \quad (9)$$

The matrix inversion is obtained recursively and the weight update equation is given by:

$$w(n) = w(n-1) + q(n)[d^*(n) - w^H x(n)] \tag{10}$$

$$q(n) = \frac{\gamma^{-1} R_{xx}^{-1}(n-1)x(n)}{1 + \gamma^{-1} x^H(n) R_{xx}^{-1}(n-1)x(n)} \tag{11}$$

$$R_{xx}^{-1}(n) = \gamma^{-1} R_{xx}^{-1}(n-1) - q(n)x(n)R_{xx}^{-1}(n-1) \tag{12}$$

The RLS algorithm is computationally more intensive, but provides much faster convergence than the LMS algorithm.

Figure 5 depicts the block diagram of Frost Beamformer or an adaptive filter-and-sum Beamformer as proposed in [14], where the filter coefficients are adapted using a constrained version of the LMS algorithm. The adaptation rule minimizes the noise power at the output while maintaining a constraint on the filter response in look direction.

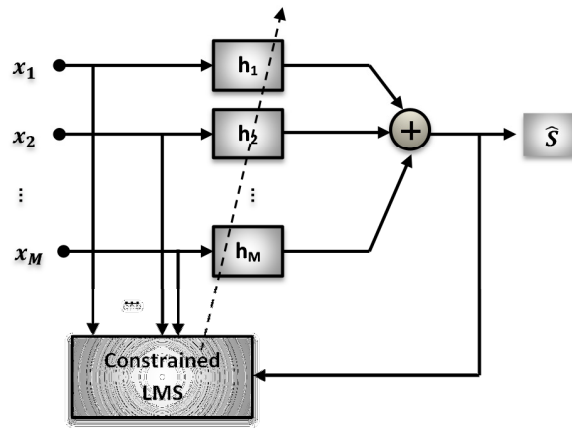

**Figure 5**. Frost Beamformer

In an MVDR Beamformer [15], the power of the output signal is minimized under the constraint that signals arriving from the assumed direction of the desired speech source are processed without distortion. An improved solution to the constrained adaptive Beamforming problem decomposes the adaptive filter-and-sum Beamformer into a fixed Beamformer and an adaptive multi-channel noise canceller. The resulting system is termed the Generalized Side-lobe Canceller [16], a block diagram of which is shown in Figure 6. Here, the constraint of a distortionless response in look direction is established by the fixed Beamformer. The noise canceller can then be adapted without a constraint.

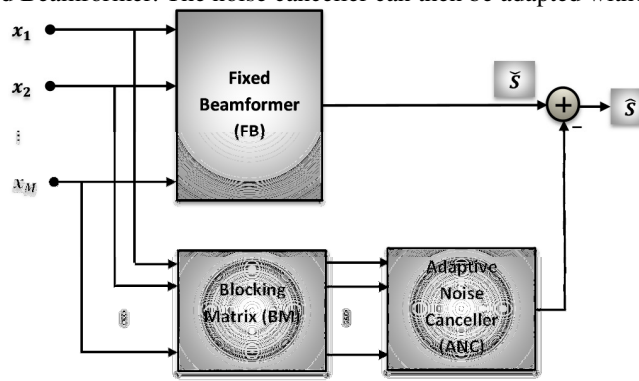

**Figure 6.** GSC Beamformer

The fixed Beamformer can be implemented via one of the previously discussed methods, for example, as a delay-and-sum Beamformer. To avoid distortions of the desired signal, the input to the adaptive noise canceller must not contain the desired signal. Therefore, a blocking matrix is employed such that the noise signals are free of the desired signal.

The adaptive noise canceller then estimates the noise components at the output of the fixed Beamformer and subtracts the estimate. Since both the fixed Beamformer and the multi-channel noise canceller might delay their respective input signals, a delay in the signal path is required. In practice, the GSC can cause a degree of distortion to the desired signal, due to a phenomenon known as signal leakage. Signal leakage occurs when the blocking matrix fails to remove the entire desired signal from the lower noise cancelling path. This can be particularly problematic for broad-band signals, such as speech, as it is difficult to ensure perfect signal cancellation across a broad frequency range. In reverberant environments, it is in general difficult to prevent the desired speech signal from leaking into the noise cancellation branch.

Several researchers have proposed modifications to the MVDR for dealing with multiple linear constraints, denoted LCMV. Their work was motivated by the desire to apply further control to the array/Beamformer beam-pattern, beyond that of a steer-direction gain constraint. Hence, the LCMV can be applied to construct a beam-pattern satisfying certain constraints for a set of directions, while minimizing the array response in all other directions.

In [10] Shmulik Markovich presented a method for source extraction based on the LCMV Beamformer. This Beamformer have the same structure of GSC, it was designed to satisfy two sets of linear constraints. One set is dedicated to maintaining the desired signals, while the other set is chosen to mitigate both the stationary and nonstationary interferences. A block diagram of this Beamformer is depicted in Figure 7:

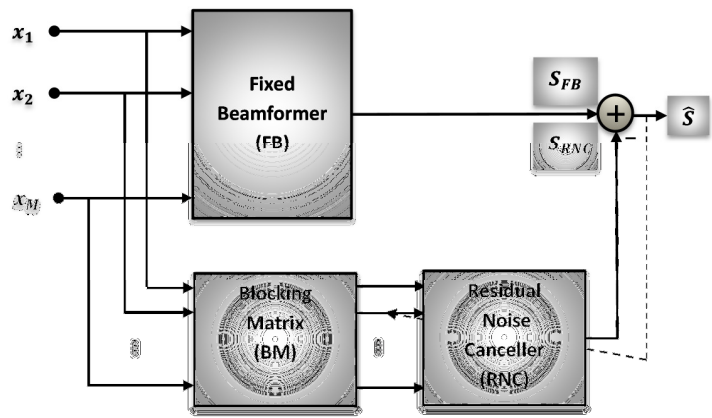

**Figure 7.** LCMV Beamformer and RNC

The LCMV Beamformer comprises three bocks the fixed Beamformer responsible for the alignment of the desired source and the blocking matrix blocks the directional signals. The output of the blocking matrix is then processed by the residual noise canceller filters for further reduction of the residual interference signals at the output. For more details regarding each block of this Beamformer and for the various definitions of the constraints see [8,10,17].

## 4. Discussion

Beamforming techniques are applied to microphone arrays in the objective of separating sources and improving intelligibility, by means of spatial filtering. The problem of source separation and extraction depends on the number of sources, the number of microphones and their arrangements, the noise level, the way the source signals are mixed within the environment, and on the prior information about the sources, microphones, and mixing parameters. Blind methods do not rely on specific characteristics of the sources, microphones, or mixing parameters. On the contrary, informed methods exploit some prior information about the sources and microphones for example, the location of a desired source. In general, the problem is more difficult when the reverberation time of the acoustic environment is large, and when the number of sources is larger than the number of microphones.

As much as the number of microphones is large, the performance of Beamforming techniques is often better. However, when the number of microphones surpassed the number of sources, perfect attenuation of all interferers becomes impossible using time-invariant Beamforming techniques and only partial interference attenuation is possible. This can be explained when in time-invariant Beamforming, the number of directions of arrival that can be perfectly cancelled is restricted by the number of microphones. In order to obtain tangible results, Beamforming techniques require information about the microphone array configuration and the sources. Few methods proposed in the literature have good separation results in a real cocktail party environment. Given the fact that almost all previous proposed methods had shortcomings, researchers resort to methods based on the combination techniques such as Beamforming and BSS [9,18,19].

## 5. Conclusion

In this paper, Beamforming techniques for multichannel signals such as speech and audio, fundamental concepts have been enhanced and recent advanced developments have only been outlined and proposed. This work is concerned with two categories for Beamforming presented as a data-independent deterministic Beamforming or data-dependent statistically optimum one.

Beamforming can enhance signals from the desired direction while hindering ones from other directions. Thus, Beamforming can be used for both noise suppression and de-reverberation. Nevertheless, its performance still degrades in cocktail-party conditions. First, the performance is closely related to the microphone array size a large array is usually required to obtain a satisfactory result but may not be practically feasible. Second, Beamforming cannot neither adequately control nor reduce reverberation coming from the desired direction. Because of the fore mentioned reasons, few methods proposed in the late works have efficient separation results in a real cocktail party environment. Actually, given these limitations, the tendency towards the use and adoption of other techniques in addition to the Beamforming techniques becomes more and more a necessary step. Indeed, researchers in this academic field became aware that the move towards the adoption of more than one type of techniques is highly useful if not recommended. Thus the adoption of the Time Frequency Masking and Independent Component Analysis.